\documentclass[pre,preprint,showpacs,showkeys,amsfonts,amsmath,amssymb,floatfix]{revtex4}

\usepackage{graphicx}
\usepackage[dvipsnames,usenames]{color}
\usepackage{dcolumn}
\usepackage{bm}
\usepackage{latexsym}
\usepackage[labelformat=empty]{subfig}
\usepackage[colorlinks]{hyperref}

\usepackage{BarnettNcomp}

\DeclareMathOperator{\EIG}		{Eig}
\DeclareMathOperator{\EMSYMM}	{\mathfrak m}

\newcommand{\gapcom}[0]			{\mathrm{\ ,}}
\newcommand{\gapstop}[0]		{\mathrm{\ .}}
\newcommand{\hfspacea}[0]		{\hspace{2.5cm}}
\newcommand{\hfspaceb}[0]		{\hspace{0.7cm}}
\newcommand{\EM}[2]				{\EMSYMM_{{#1},{#2}}}
\newcommand{\EMS}[0]			{\EMSYMM_1}
\newcommand{\erdren}[0]			{{Erd\"os-R\'enyi}}
\newcommand{\dt}[0]				{\,dt}
\newcommand{\beps}[0]			{\boldsymbol\varepsilon}
\newcommand{\bX}[0]				{\mathbf{X}}
\newcommand{\bW}[0]				{\mathbf{W}}
\newcommand{\cI}[0]				{\mathcal{I}}
\newcommand{\ND}[0]				{\mathcal{N}}
\newcommand{\CN}[0]				{\mathcal{C_N}}
\newcommand{\CNA}[0]			{\mathcal{C}^*_N}
\newcommand{\CNAA}[0]			{\mathcal{C}^{**}_N}
\newcommand{\Expect}[1]			{\braca{#1}}

\begin{document}


\title{A Graph Theoretic Interpretation of Neural Complexity}

\author{L. Barnett}
\email{l.c.barnett@sussex.ac.uk}

\affiliation{
	Neurodynamics and Consciousness Laboratory and \\
	Sackler Centre for Consciousness Science \\
	School of Informatics \\
	University of Sussex, BN1 9QH, UK
}

\author{C. L. Buckley}
\email{c.l.buckley@sussex.ac.uk}

\affiliation{
	Centre for Computational Neuroscience and Robotics \\
	School of Informatics \\
	University of Sussex, BN1 9QH, UK
}

\author{S. Bullock}
\email{sgb@ecs.soton.ac.uk}

\affiliation{
	School of Electronics and Computer Science \\
	University of Southampton, SO17 1BJ, UK
}

\date{\today}

\begin{abstract}

One of the central challenges facing modern neuroscience is to explain the ability of the nervous system to coherently integrate information across distinct functional modules in the absence of a central executive. To this end Tononi \etal\ [Proc. Nat. Acad. Sci. USA \textbf{91}, 5033 (1994)] proposed a measure of \emph{neural complexity} that purports to capture this property based on mutual information between complementary subsets of a system. Neural complexity, so defined, is one of a family of information theoretic metrics developed to measure the balance between the segregation and integration of a system's dynamics.  One key question arising for such measures involves understanding how they are influenced by network topology. Sporns \etal\ [Cereb. Cortex \textbf{10}, 127 (2000)] employed numerical models in order to determine the dependence of neural complexity on the topological features of a network. However,  a complete picture has yet to be established. While De Lucia \etal\ [Phys. Rev. E \textbf{71}, 016114 (2005)] made the first attempts at an analytical account of this relationship, their work utilized a formulation of neural complexity that, we argue, did not reflect the intuitions of the original work.

In this paper we start by describing weighted connection matrices formed by applying a random continuous weight distribution to binary adjacency matrices. This allows us to derive an approximation for neural complexity in terms of the moments of the weight distribution and elementary graph motifs. In particular we explicitly establish a dependency of neural complexity on cyclic graph motifs.

\end{abstract}



\maketitle

\section{Introduction} \label{sec:intro}

Graph theory has been employed with some success within neuroscience \cite{bab06}. However graph theory explicitly concerns itself with characterizing network structure and at best can deliver only circumstantial insight into network dynamics. As a partial answer to this, over the past decade neuroscientists  have developed a suite of dynamical measures that employ  information theory \cite{tse94,t03,s05}. While application of information theoretic measures come with their own limiting assumptions (\eg\ they typically demand that the dynamics on a network be stationary)  their ability to deal naturally with weighted connectivity matrices has made them extremely appealing. However in general the analytical understanding of such measures is less well-developed than the tools of graph theory. Consequently,  it is likely that  a more comprehensive  picture of network function could be achieved by an understanding of the  relationship between information theoretic  and graph theoretic measures \cite{stk05}. In this paper we pursue one such analysis.

Neural complexity, as formulated by Tononi, Sporns and Edeleman \cite{tse94}, is perhaps one of the most widely discussed information theoretic measures. It purports to capture the nervous system's ability to coherently integrate information at the level of the organism  while maintaining  the functional segregation associated with specialization across a range of anatomical scales \cite{tes98}.

The TSE measure calculates the mutual information shared between processes occurring on complementary subsystems of a neural network in order to identify networks that simultaneously support relatively segregated local behavior and relatively integrated global dynamics. Tononi \etal\ were able to demonstrate that network structures inspired by the properties of the cerebral cortex scored highly \cite{tse94,ste00a}. Subsequently, the measure has been used extensively, \eg\ to characterize the dynamics of different regions in the mammalian brain \cite{ste00b}, to analyze evolved robot controllers \cite{ase04} and to explore theories of sleep, consciousness and schizophrenia \cite{te98}.

The relationship between neural complexity and graph topology has been central to the intuitions that underlie the neural complexity  measure. Originally Sporns \etal\ \cite{ste00b} found that constructing the graph topology of networks using organizational principles derived from observations of the cerebral cortex led to high neural complexity. Sporns \etal\ \cite{ste00a} used graph theoretic tools to analyze the topology of networks  that were optimized for high complexity using an evolutionary algorithm.  De Lucia \etal\ \cite{del05}  were  the first to attempt to decompose neural complexity in terms of elementary graph  motifs. They achieved this by utilizing a popular analytic model that proceeds on the assumption that neural dynamics may be approximated by a stationary multivariate stochastic process and, furthermore, that this process is Gaussian. This enables the interactions between network components (and mutual information itself) to be expressed via a covariance matrix.  Recently we have highlighted and then amended an error in this analytical model by moving to a continuous time analogue of the original apparently discrete time formulation \cite{bbb08a}. Consequently it is necessary to revisit the analytical work of De Lucia \etal\ in light of this correction. Furthermore, we argue that the formulation of neural complexity employed by De Lucia \etal\ did not implement correctly the original definition of neural complexity \cite{bbb08a}.

We start  by deriving an approximation for neural complexity in terms of elementary graph motifs. We find that to a first approximation  complexity is dependent on reciprocal connections and to a second approximation on two types of 3-cycle. We go on to suggest that higher order terms in our approximation for neural complexity will be dependent, with diminishing impact, on cycles of increasing order.

We illustrate the  validity of our approximation by example of a  ring lattice connectivity scheme where the probability of connection between elements decays with the distance between them. This scheme may be considered a discretized version of the model discussed in detail in \cite{bbb08a}, which was inspired by a Toeplitz covariance matrix model suggested in \cite{tse94}. The ring lattice model ratifies our application of the neural complexity approximation to topological graphs and also reveals conditions for the presence of a peak in complexity at intermediate connectivity decay values.

\section{Neural complexity} \label{sec:ncomp}

The scenario we address here is the same as in \cite{bbb08a} (see also \cite{tse94, tse96, tse99, ste00a}): we have a system of $n$ ``neural components'' (\emph{nodes} for brevity) and a stationary multivariate stochastic  process $\bX(t) \equiv \{X_i(t) | \ i = 1,\ldots,n\}$ running on the system, where $X_i(t)$ represents the activation state at time $t$ of the $i^{\mathrm{th}}$ node. In \cite{tse94} the authors introduced a \emph{neural complexity} measure based on mutual information between subsystems of the given system. The idea behind the measure is that complex neural systems should be expected to exhibit a balance between ``integration'' and ``segregation'' of neural subsystems. The measure is defined as follows: firstly the \emph{integration} associated with the system is introduced as
\begin{equation}
	\cI \equiv \sum_{i=1}^n H_i - H \gapcom
\end{equation}
where $H$ denotes the entropy $ \entro{\bX(t)}$ of the full process $\bX(t)$ and $H_i$ the entropy $\entro{X_i(t)}$ of the individual activation $X_i(t)$. Note that by stationarity these quantities and hence $\cI$ itself do not depend on time $t$. $\cI$ may be interpreted as a measure of the deviation from independence of the individual components of the system. Neural complexity is then defined to be
\begin{equation}
	\CN \equiv \sum_{k=1}^{n-1} \bracr{\frac k n \cI - \braca \cI_k} \gapcom \label{eq:ncomp1}\
\end{equation}
where $\braca\cdot_k$ denotes an average \emph{over all subsystems of size $k$} of the given system. Neural complexity is thus an average over all scales (represented by subsystem size $k$) of the difference between mean integration $\braca \cI_k$ at the given scale and the appropriately scaled global integration $\frac k n \cI$. For a highly segregated system both of these quantities will be small and $\CN$ itself thus small. Conversely, for a highly integrated system the integration for individual subsystems will be close to the scaled global integration, and $\CN$ will again be small. $\CN$ can thus be expected to attain peak values for systems which are neither highly integrated nor highly segregated; see \cite{tse94} for a fuller discussion of interpretation of the measure. $\CN$ may also be expressed directly in terms of entropies as
\begin{equation}
	\CN = \sum_{k=1}^{n-1} \bracr{\braca H_k -\frac k n H} \gapstop \label{eq:ncomp}
\end{equation}

\subsection{Actualization of the measure} \label{sec:ncompact}

In the special case where the $\bX(t)$ are jointly \emph{multivariate Gaussian}, the entropy $H$ may be expressed simply in terms of the $n \times n$ covariance matrix $\Omega \equiv \overline{\trans{\bX(t)} \bX(t)}$, where the over-bar represents an average over the statistical ensemble  \cite{infot}. By stationarity $\Omega$ does not depend on time $t$. We then have $H = \shalf \ln\bracr{\bracs{2\pi e}^n \abs{\Omega}}$ so that
\begin{equation}
	\CN = \half \sum_{k=1}^{n-1} \bracr{\braca{\ln{\abs\Omega}}_k -\frac k n \ln{\abs\Omega}} \gapstop \label{eq:ncompgd}
\end{equation}

Tononi \etal\ \cite{tse94} consider an $n \times n$ \emph{connectivity matrix} $C$, where $C_{ij}$ is to be interpreted as the \emph{weight} on the connection from node $i$ (efferent) to node $j$ (afferent), and a linear autoregressive neural process
\begin{equation}
	\bX(t) = \bX(t) \cdot C + \beps(t) \label{eq:procd}
\end{equation}
driven by serially uncorrelated Gaussian noise $\beps(t)$. However, there is an error in their calculation of the covariance matrix associated with the process \eqref{eq:procd} \cite{bbb08a}. While the error is readily corrected, in \cite{bbb08a} it is argued that such a \emph{discrete time} process is likely to be unacceptably unrealistic and leads, furthermore, to conclusions which probably do not support the intuitions of the originators of the $\CN$ measure \cite{tse94}. Thus in \cite{bbb08a} the discrete time model \eqref{eq:procd} is dropped in favor of the continuous time multivariate \emph{Ornstein-Uhlenbeck} process \cite{uhlorn30,oksen}
\begin{equation}
	d\bX(t) = -\bX(t) \cdot (I-C)\dt + d\bW(t) \gapcom \label{eq:procou}
\end{equation}
where $\bW(t)$ is a multivariate Wiener process with identity covariance matrix, representing white noise applied independently to each node \footnote{Note that the noise input to different nodes is uncorrelated. If we allow noise \emph{levels} to differ per node, then we may recover an equivalent equation to \eqref{eq:procou} by a simple linear transformation of the connectivity matrix and a rescaling of activation levels.}. In \cite{bbb08a} it is shown that neural complexity for this process has distinctly different characteristics from that for the discrete time process \eqref{eq:procd} (see \cite{bbb08a} for more discussion on this topic).

Eq.~\eqref{eq:procou} may be viewed as a linearized, noisy Continuous Time Recurrent Neural Network (CTRNN) \cite{gal08}. The condition for existence of a stationary process \eqref{eq:procou} is
\begin{equation}
	\max\bracc{\rcond{\re(\lambda)}{\lambda \in\EIG(C)}} < 1 \gapcom \label{eq:stat}
\end{equation}
where $\EIG(C)$ denotes the set of eigenvalues of $C$. The stationary process \eqref{eq:procou} is multivariate Gaussian so that \eqref{eq:ncompgd} applies, with covariance matrix $\Omega$ satisfying the continuous-time Lyapunov equation
\begin{equation}
	2\Omega = I + \trans C \Omega + \Omega C \gapcom \label{eq:cov}
\end{equation}
(see \cite{bbb08a} for a derivation) for which there exist efficient algorithms for numerical solution \cite{barstew72}.

\subsection{The neural complexity approximation} \label{sec:ncompx}

From \eqref{eq:cov} the matrix series expansion
\begin{equation} \begin{split} \label{eq:omsx}
	2\Omega
	= & \ \sum_{r=0}^\infty 2^{-r} \sum_{k = 0}^r \binom r k \trans{\bracr{C^k}} C^{r-k} \\
	= & \ I + \tfrac 1 2 \bracr{\trans C + C} \\
	& \ \phantom{I} + \tfrac 1 4 \bracs{{\trans{\bracr{C^2}} + 2 \trans C C + C^2}} + \cdots
\end{split} \end{equation}
may be derived. Defining the order parameter $\epsilon \equiv \dabs C$ for some (submultiplicative) matrix norm $\dabs\cdot$ \cite{ma}, in \cite{bbb08a} this expansion is the basis for derivation of the approximation
\begin{equation}
	\CN = \CNA + \CNAA + \bigO{\epsilon^4} \gapcom \label{eq:cn}
\end{equation}
where
\begin{equation}
	\CNA \equiv \phantom{+} \frac{n+1}{48} \,\, \sum_{i \ne j} \bracr{\subsa C{ij}2 + C_{ij} C_{ji}} \label{eq:ccna}
\end{equation}
is $\bigO{\epsilon^2}$ and
\begin{equation} \begin{split} \label{eq:ccnaa}
	\CNAA \equiv & \phantom+ \ \frac{n+1}{96} \sum_{i \ne j \ne k} \bracr{3 C_{ij} C_{jk} C_{ik} + C_{ij} C_{jk} C_{ki}} \\
		  & + \frac{n+1}{24} \,\, \sum_{i \ne j} C_{ii} \bracr{\subsa C{ij}2 + C_{ij} C_{ji}}
\end{split} \end{equation}
is $\bigO{\epsilon^3}$.

In \cite{bbb08a} it is recommended that, in order to establish a level playing field when comparing complexity between networks, some form of \emph{normalization} be applied to the connection matrix $C$; in particular \emph{spectral normalization}, where $C$ is premultiplied by $w/\rho(C)$ with
\begin{equation}
	\rho(C) \equiv \max\bracc{\rcond{\abs\lambda}{\lambda \in\EIG(C)}} \label{eq:srad}
\end{equation}
the \emph{spectral radius} of $C$ and $0 < w < 1$ a scale parameter. Since spectral radius is the infimum of all (induced) matrix norms \cite{ma} it is a good indicator of the accuracy of the neural complexity approximation \eqref{eq:cn}; \ie\ the approximation can be expected to be accurate for small $w$. This is borne out empirically \cite{bbb08a}.

\section{Relating network connectivity to graph structure} \label{sec:nc2gs}

The motivation for this paper is to investigate how neural complexity $\CN$ relates to the \emph{graph structure} of a putative network underlying a neural system. Towards this end we require some plausible scheme by which to relate a connectivity matrix $C$ to a given adjacency matrix $A \equiv \bracr{A_{ij}}$ representing the topology of the underlying network.

De Lucia \etal\ \cite{del05} introduce a simplified scheme where the connection matrix $C$ is just the adjacency matrix, which they take to be symmetric and with no self-connections, normalized by system size. They then proceed to derive an approximation to neural complexity $\CN$ for a stationary multivariate Gaussian process on such a network. However, besides taking as their starting point the discrete time process \eqref{eq:procd} \footnote{We remark that the previously mentioned covariance calculation error of \cite{tse94} (see Section~\ref{sec:ncomp}) is not repeated in \cite{del05}; covariance matrices are (implicitly) calculated correctly according to their eq.~3.} which, as mentioned previously, we consider unsuited to neural complexity analysis, we believe their analysis to be flawed in the following respect: in averaging over $k$-subsystems as required by \eqref{eq:ncomp}, the authors of \cite{del05} appear effectively to treat  subsystems \emph{in isolation}---\ie\ as neural processes in their own right, uncoupled from the full system \footnote{See \eg\ the derivation in \cite{del05} of their eq.~14, where the quantity $D_2(k)$  seems to denote the quantity $D_2$ previously introduced for the full system of $n$ nodes, but interpreted for an \emph{independent} system of $k$ nodes.}---which surely defeats the purpose of $\CN$ as defined by \eqref{eq:ncomp1}. This has the result of introducing spurious $2^{\mathrm{nd}}$ order terms (the correct result for the discrete system time system is of $4^{\mathrm{th}}$ order in connectivity; see \cite{bbb08a}, eq.~40).

In this study we consider only \emph{directed} graphs. Our approach, however, extends straightforwardly to \emph{undirected} (bi-directional) graphs. Graph topology is specified by a binary adjacency matrix $A_{ij}$, so that $A_{ij} = 1$ represents a directed connection from node $i$ (efferent) to node $j$ (afferent). As regards diagonal elements, we take the view here that, while our graphs will not have self-connections---so that $A_{ii} = 0$ for all $i$---diagonal elements of the \emph{connection} matrix are to be regarded as representing variation in the characteristic relaxation time, or activation decay, of the individual neural components. See \cite{bbb08a} for more discussion on this point. In the spirit of statistical physics we now suppose that, given an adjacency matrix $A$, weights are assigned \emph{independently and identically at random} to each connection. We suppose that diagonal activation decay elements are drawn, also identically and independently, from a separate distribution with zero mean, so that
\begin{equation} \label{eq:atoc}
	C_{ij} = \left\{ \begin{array}{ccc}
		W_{ij} A_{ij} && i \ne j \\
		&& \\
		D_i && i = j
	\end{array} \right.
\end{equation}
where the $W_{ij}$ are iid as some random variable $W$ and the $D_i$ iid as some random variable $D$ which has mean zero. We may then, given a graph structure $A$, consider neural complexity $\CN$---via the randomness \footnote{But note that the variances of $W$ and $D$ may be zero in the degenerate case of there being no randomness.} introduced by $W,D$---as a \emph{random variable} $\CN\!\bracr{A;W,D}$ and we \emph{define} the complexity measure for a graph $A$ to be the expectation over $W, D$:
\begin{equation}
	\CN(A) \equiv \expect{\CN\!\bracr{A;W,D}} \gapstop \label{eq:cnmean}
\end{equation}
We define similarly the approximations $\CNA(A)$, $\CNAA(A)$. Note that these measures depend on the particular distributions $W,D$ used in the construction of the random connectivity matrix. As regards normalization, in the case where graphs of similar \emph{mean degree} are compared one might take the view that normalization is less critical, since in some sense the overall connectivity strength averaged over all network weights $W,D$ will not vary drastically (but see the example in Section~\ref{sec:sworld} below).

\begin{figure*}
\begin{center}
$\begin{array}{c@{\hfspacea}c@{\hfspacea}c@{\hfspacea}c}
	\subfloat[$\EM 2  1 $]{\includegraphics{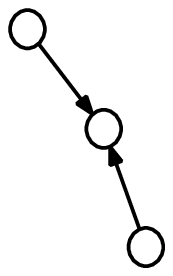}} &
	\subfloat[$\EM 2  2 $]{\includegraphics{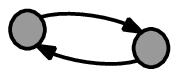}} &
	\subfloat[$\EM 2  3 $]{\includegraphics{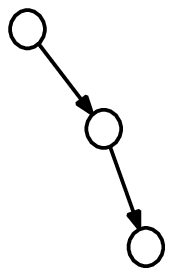}} &
	\subfloat[$\EM 2  4 $]{\includegraphics{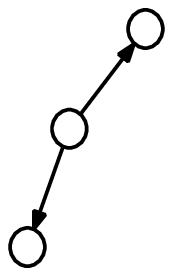}}
\end{array}$
\caption{$2$-edge motifs: only shaded motifs are implicated in the neural complexity measure.} \label{fig:motifs2}
\end{center}
\end{figure*}

\begin{figure*}
\begin{center}
$\begin{array}{c@{\hfspaceb}c@{\hfspaceb}c@{\hfspaceb}c}
	\subfloat[$\EM 3  1 $]{\includegraphics{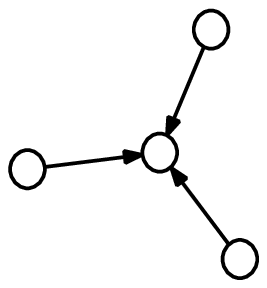}} &
	\subfloat[$\EM 3  2 $]{\includegraphics{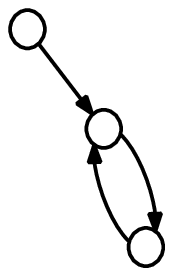}} &
	\subfloat[$\EM 3  3 $]{\includegraphics{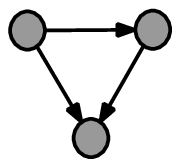}} &
	\subfloat[$\EM 3  4 $]{\includegraphics{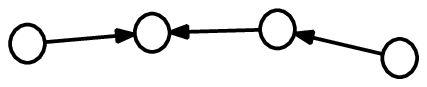}} \\
	\subfloat[$\EM 3  5 $]{\includegraphics{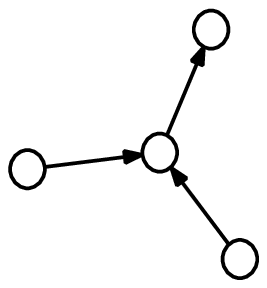}} &
	\subfloat[$\EM 3  6 $]{\includegraphics{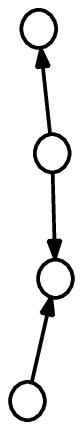}} &
	\subfloat[$\EM 3  7 $]{\includegraphics{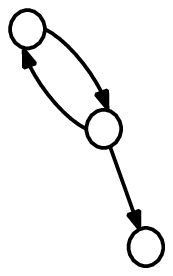}} &
	\subfloat[$\EM 3  8 $]{\includegraphics{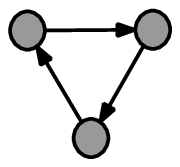}} \\
	\subfloat[$\EM 3  9 $]{\includegraphics{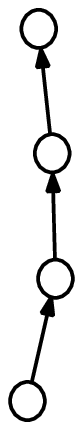}}  &
	\subfloat[$\EM 3{10}$]{\includegraphics{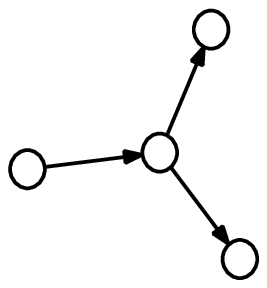}} &
	\subfloat[$\EM 3{11}$]{\includegraphics{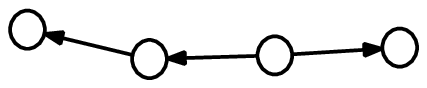}} &
	\subfloat[$\EM 3{12}$]{\includegraphics{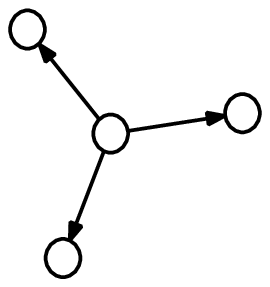}}
\end{array}$
\caption{$3$-edge motifs: only shaded motifs are implicated in the neural complexity measure.} \label{fig:motifs3}
\end{center}
\end{figure*}

Without normalization---and assuming that we may safely ignore instances of $C$ for which the stationarity condition \eqref{eq:stat} fails---we may calculate from \eqref{eq:ccna} and \eqref{eq:ccnaa} that
\begin{eqnarray}
	\CNA(A)  & = & \frac{n+1}{48} \bracr{\mu_2 \EMS + 2 \mu^2 \EM 2 2} \label{eq:eccna} \\
	\CNAA(A) & = & \frac{n+1}{32} \mu^3 \bracr{\EM 3 3 + \EM 3 8} \gapcom \label{eq:eccnaa}
\end{eqnarray}
where $\mu \equiv \expect W$ and $\mu_2 \equiv \expect{W^2}$ are respectively the mean and second moment of the weight distribution, and
\begin{eqnarray}
	\EMS    & \equiv & \phantom{\tfrac 1 2}  \sum_{i,j} A_{ij} \\
	\EM 2 2 & \equiv & \shalf \sum_{i,j} A_{ij}A_{ji} \label{eq:R} \\
	\EM 3 3 & \equiv & \phantom{\tfrac 1 2} \sum_{i,j,k} A_{ij} A_{jk} A_{ik} \label{eq:L1} \\
	\EM 3 8 & \equiv & \tfrac 1 3 \sum_{i,j,k} A_{ij} A_{jk} A_{ki} \gapstop \label{eq:L2}
\end{eqnarray}
The $\EM p q$ may be interpreted as the multiplicities of certain \emph{graph motifs}---small repeated subgraph fragments, \cite{msikca02}, in the graph represented by $A$. Specifically, $\EMS$ is just the total number of edges of the graph, $\EM 2 2$ counts the number of \emph{reciprocal} connections, while $\EM 3 3$ and  $\EM 3 8$ count the numbers of two varieties of $3$-cycle; see FIGS.~\ref{fig:motifs2}~and~\ref{fig:motifs3}.
Note that since activation decay $D$ has zero mean neither $\CNA(A)$ nor $\CNAA(A)$ depends on $D$; there may potentially, however, be dependencies on higher order moments of $D$ for higher-order approximations.

Equations \eqref{eq:eccna} and \eqref{eq:eccnaa} tell us the following: the expected neural complexity of a neural system based on a graph according to the prescription \eqref{eq:atoc} depends, to a first approximation, on the total number of connections and the number of reciprocal connections in the graph. To the next order of approximation it depends on the number of two varieties of $3$-cycle. The procedure via which $\CNA$ and $\CNAA$ were derived (see \cite{bbb08a}) suggests that if we were to calculate higher order approximations according to the expansion \eqref{eq:omsx} for the covariance matrix, then higher order cyclic graph motifs would successively come into effect with diminishing impact, to a degree depending roughly on the spectral radius of $A$ and the moments of $W$ (and possibly of $D$).

We may well wish to assess neural complexity for \emph{statistical ensembles} of graph structure; that is, the adjacency matrix elements $A_{ij}$ represent (jointly distributed) random variables. In this case, $\CN(A)$ and its approximations, as well as the motif counts $\EMS,\EM 2 2,\EM 3 3,\EM 3 8$  may themselves be considered as random variables and their ensemble means (and higher moments) calculated. In the following we shall use the notation $\Expect\cdot$ to denote an average over a statistical ensemble of graphs, as distinct from the mean $\expect\cdot$ over the network weight random variables $W, D$.

\section{Example} \label{sec:sworld}

To illustrate the integration/segregation balance that motivates their formulation of the measure $\CN$, \cite{tse94} demonstrate a complexity peak for an \textit{ad hoc} Toeplitz covariance matrix where covariance decays as we move away from the diagonal. In \cite{bbb08a} this idea is analyzed in more detail via our approximation formula. In that paper a connection matrix is constructed based on a ring lattice, where connectivity (and consequently covariance) decays exponentially with lattice distance (see also \cite{bb07}). The resultant covariance matrix is a Toeplitz matrix analogous to that in \cite{tse94}. Here we introduce a model which parallels the ring lattice construction in \cite{bbb08a}. Rather than, as in that model, having connection \emph{strength} decay with lattice distance, we instead construct an ensemble of random graphs where the \emph{probability} of connection decays with lattice distance. This model is thus amenable to a graph theoretic analysis as expounded above.

We proceed by constructing a directed random graph with no self-connections on a ring lattice of $n$ nodes, where each edge is assigned independently with probability
\begin{equation}
	\prob{A_{ij} = 1} = P_{ij} \equiv c a^{d(i,j)} \label{eq:swaij}
\end{equation}
for $i \ne j$, where $d(i,j) \equiv \min\bracr{\abs{i-j},n-\abs{i-j}}$ is ring lattice distance, $0 < a < 1$ a connectivity decay parameter \footnote{The connectivity decay parameter $a$ is somewhat akin to the ``rewiring'' parameter in small world models \cite{w99}, to which our model bears some resemblance.} and $c$ a constant to be determined. We shall generally work to the large network limit $n \rightarrow \infty$. We may calculate that in this limit the \emph{mean (in or out) degree} is given by $\kappa = 2cr/(1-a)$. We shall take $n$,  $\kappa$ and $a$ as the defining parameters of the model, so we set
\begin{equation}
	c \equiv \shalf \kappa \bracr{1/a-1} \label{eq:cee}
\end{equation}
in \eqref{eq:swaij}. Note that $P_{ij} \le 1$ for all $i \ne j$ requires
\begin{equation}
	a > a_0 \equiv \min(1-2/\kappa,0) \gapstop \label{eq:arr}
\end{equation}
The probability that two arbitrarily chosen distinct nodes are connected is given by $p \equiv \kappa/n$; it corresponds to the connection probability for an \erdren\ random graph \cite{b01} of the same size. In the large network limit, the degree distribution of our model tends to a Poisson distribution, as for an \erdren\ random graph \cite{b01}. Note that while the mean degree of individual graphs instantiated from the ensemble may vary, its ensemble average is always $\kappa$.

We first calculate the mean approximate neural complexity without normalization; to this end the chief advantage of our model lies in the mutual independence of the $A_{ij}$ which allows us to calculate easily the means of the quantities $\EMS, \EM 2 2, \EM 3 3, \EM 3 8$ over the ensemble. In the large network limit we find
\begin{eqnarray}
	\Expect{\EMS}   & = & n\kappa \label{eq:EK} \\
	\Expect{\EM 2 2}   & = & \tfrac 1 4 n\kappa^2 \frac{1-a}{1+a} \label{eq:ER} \\
	\Expect{\EM 3 3} = 3\Expect{\EM 3 8} & = & \tfrac 3 4 n\kappa^3 \frac{a(1-a)}{(1+a)^2} \gapcom \label{eq:EL1L2}
\end{eqnarray}
valid for $a_0 < a < 1$. Recall that in \eqref{eq:eccna} and \eqref{eq:eccnaa} $\CNA(A)$ and $\CNAA(A)$ have already been averaged over the weight distribution $W$ applied to a given graph specified by $A$. For fixed $n, \kappa, a$ we may now, using \eqref{eq:EK} - \eqref{eq:EL1L2}, average again over the ensemble of random graphs ${A(n,\kappa,a)}$ to derive the mean neural complexity approximations $\Expect{\CNA(n,\kappa,a)}$ and $\Expect{\CNAA(n,\kappa,a)}$. We find
\begin{eqnarray}
	\Expect{\CNA(n,\kappa,a)}  & = & \frac{n(n+1)}{48} \bracr{\kappa \mu_2 + \shalf \kappa^2 \mu^2 \frac{1-a}{1+a}} \label{eq:eccnasw} \\
	\Expect{\CNAA(n,\kappa,a)} & = & \frac{n(n+1)}{32} \kappa^3 \mu^3 \frac{a(1-a)}{(1+a)^2} \gapstop \label{eq:eccnaasw}
\end{eqnarray}
For fixed $n,\kappa$ the first term in $\Expect{\CNA(n,\kappa,a)}$ is constant, while the second term is monotone decreasing with $a$. $\Expect{\CNAA(n,\kappa,a)}$ has a maximum at $a = \tfrac 1 3$; rewiring initially boosts and then degrades the number of $3$-cycles. To examine the behavior of the unnormalized approximation in more detail, we may calculate that, as a function of $a$, $\Expect{\CNA(n,\kappa,a)}+\Expect{\CNAA(n,\kappa,a)}$ has an extreme value at
\begin{equation}
	a^* = \frac{3\kappa\mu-2}{9\kappa\mu+2} \gapcom
\end{equation}
although the extremal only represents a \emph{maximum} (\ie\ a complexity peak) if $\mu < 0$, which implies a predominance of \emph{inhibitory} connections. We note that while this condition may exclude the mammalian cortex, where the majority of connections are thought to be excitatory, it may be relevant to early sensory relays which are dominated by inhibition \eg\ the olfactory bulb or the first stages of the visual system \cite{shepherd2003}.
In any case, there exists the possibility of a complexity peak for negative $\mu$; then the condition $a_0 < a^* <1$ implies
\begin{align}
	1 < \kappa \le 3 &&\textrm{ and }&& \mu < -\frac 2{3\kappa}  \\
	\textrm{or} \nonumber \\
	\kappa > 3 &&\textrm{ and }&& -\frac 2{3\kappa} \bracr{\frac{\kappa-1}{\kappa-3}} \le \mu < -\frac 2{3\kappa}
\end{align}
(bearing in mind that if $\mu$ is too large and negative then the stability condition \eqref{eq:stat} is likely to be violated). Outside of this regime $\Expect{\CNA(n,\kappa,a)+\CNAA(n,\kappa,a)}$ is always monotone decreasing with $a$: the complexity contribution of reciprocal connections overwhelms, on average, the contribution of $3$-cycles and there is no complexity peak at intermediate connectivity decay. We remark that for a corresponding \emph{undirected} graph model every link is effectively reciprocal and there is then a possibility of a complexity peak, albeit for sparsely-connected networks, specifically for mean degree $\kappa < 3$.

We now examine neural complexity under spectral normalization. Since this case appears intractable to analysis, we carried out simulations as follows: for a series of mean degree values $\kappa$ ranging from $2.5$ to $5.5$ and for a sequence of decay parameters $a$ in the range $(a_0,1)$ we generated $10^5$ random directed graphs of $n = 30$ nodes, according to the prescription of \eqref{eq:swaij}. For each graph $A$, connectivity coefficients $C$ were generated according to \eqref{eq:atoc}. There was no decay variance; \ie\ the diagonal weights $D_i$ were set to zero. Inter-node weights $W_{ij}$ were drawn from a binormal distribution including both positive (excitatory) and a smaller fraction of negative (inhibitory) weights; specifically, each weight was drawn independently with $80\%$ probability from $\ND(0.5,0.01)$ and with $20\%$ probability from $\ND(-0.4,0.01)$. Spectral normalization was then applied to each resulting connectivity matrix with scale parameter $w = 0.2$ (we verified that at this normalization level $\CNA+\CNAA$ was generally a close approximation to the exact value $\CN$). Finally, the means $\Expect{\CNA(n,\kappa,a)}$ and $\Expect{\CNAA(n,\kappa,a)}$ were calculated in sample (given the large sample size standard errors for the estimated mean were small, although the variance was substantial).

\begin{figure}
\begin{center}
	\includegraphics{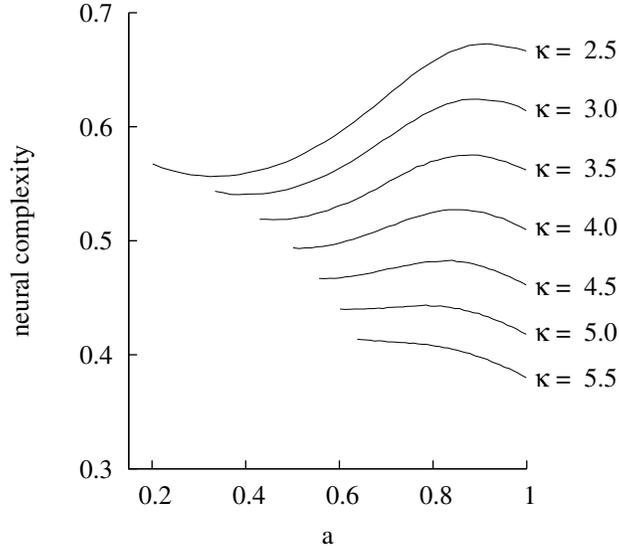}
\end{center}
\caption{Mean approximate neural complexity $\Expect{\CNA(n,\kappa,a)} + \Expect{\CNAA(n,\kappa,a)}$ under spectral normalization, plotted against connectivity decay $a$ in $30$-node networks for a range of mean degree values $\kappa$. The cut-off at the left of the plots marks the minimum connectivity decay $a_0 = 1-2/\kappa$. See main text for details.} \label{fig:xdec}
\end{figure}

Results are illustrated in FIG.~\ref{fig:xdec}.
We see that within the illustrated mean degree range there is a connectivity peak which disappears at approximately $\kappa > 5$. Thus, as long as connectivity is not too dense, we can expect a connectivity peak at intermediate connectivity decay comparable with that seen in the continuous exponential connectivity decay model of \cite{bbb08a} under spectral normalization. As in that study, normalization was (at least for positive mean connection strength $\mu$) a prerequisite for the appearance of a connectivity peak, although in the current model no self-activation was necessary.

\section{Discussion} \label{sec:outro}

The central claim of this study is that neural complexity is driven by the presence of specific topological features of a network, namely cyclic motifs, with lower-order cycles exerting the greatest influence. While here we carry the topological analysis up to third order in connectivity, our analytical approach points to the emergence of higher-order cycles with diminishing influence on neural complexity \cite{bbb08a}.

The original measure, defined for an arbitrary (stationary) stochastic process, is commonly characterized in terms of a multivariate process running on a weighted network. Consequently, in addition to specifying a network topology, one must specify weights over the network's edges before the measure can be applied. Here we chose a statistical mechanics approach where the weights on a graph are drawn from a random distribution. This allows us to consider the neural complexity associated with a particular topology by determining the mean over the ensemble of weighted networks with which it is associated.

It is instructive to contrast the behavior of the neural complexity measure for the ring-lattice model with exponential connectivity decay described in \cite{bbb08a}, where connection \emph{strength} decayed with lattice distance, and the behavior of the same measure for the model explored here in which connection \emph{probability} decays with lattice distance. The original model demonstrated that without normalization neural complexity decreases monotonically with the decay parameter, whereas with normalization there may be an intermediate complexity peak provided that there is some variance in the time constants of individual nodes. In the current model, the influence of the decay parameter on neural complexity is seen explicitly to derive from its impact on the relative prevalence of reciprocal connections and 3-cycles. While the impact of reciprocal connections simply decreases monotonically with the decay parameter, that of 3-cycles peaks at an intermediate value. How this balance plays out is determined by a network's mean degree and mean connection strength.

In general, for there to be an intermediate peak in complexity, mean degree must be low in order to prevent reciprocal connections from dominating. In addition, for directed graphs, the mean connection strength must be negative, implying a predominance of inhibitory connections. We have already mentioned that this constraint may only be met by some kinds of neural system \cite{shepherd2003}. For undirected graphs we note that the number of reciprocal connections is by definition constant, but low mean degree is still required in order to see an intermediate peak in complexity. In contrast to our previous lattice model with decaying connection strength \cite{bbb08a}, here it was not necessary to introduce variability in network node time constants.

We have seen that connectivity normalization can have a significant impact on the behavior of the neural complexity measure. Spectral normalization, for instance, changes the regime within which an intermediate complexity peak may appear, specifically accommodating higher mean degree.

In summary, while the founding intuition that complexity derives from a balance between integration and segregation might lead one to expect that an intermediate complexity peak should be a robust property of any model in which a ``segregated'' lattice-like graph is relaxed towards a more ``integrated'' random graph, we find that the picture is more subtle and that relative frequencies of specific cyclic topological features play a crucial role.

Finally, we note that neural complexity may be considered as just one of a family of information theoretic metrics based on the intuition of segregation-integration balance; we expect that the methods employed in this paper can be readily extended to related measures. Particularly promising are measures such as \emph{causal density} \cite{s05,asi06,bbs10} which, in contrast to the ``static'' mutual information underlying $\CN$, take into account \emph{directed} information flow between network subsystems, expressible in terms of \emph{transfer entropy} \cite{schreiber00,bbs09}. In future studies we intend to extend the analytic techniques presented here to such measures.

\section{Acknowledgments}

The authors would like to thank Anil Seth and Annamaria Cucinotta for useful discussions. This study was partially supported by the Spatially Embedded Complex Systems Engineering (SECSE) project, EPSRC (UK) grant no. EP/C51632X/1 and BBSRC (UK) grant no. BB/F005113/1.

\bibliography{BarnettNcomp}

\end{document}